\documentclass{icrc29}
\usepackage{graphicx,amssymb,amsmath,times,lineno}
\begin{document}

\title[Quality studies of the data taking conditions for the Auger
Fluorescence Detector]{Quality studies of the data taking conditions for the Auger
Fluorescence Detector}
\author[R. Caruso et al.]{R. Caruso,  R. Fonte, A. Insolia, 
S. Petrera and J. Rodriguez Martino 
\newauthor
for the Auger Collaboration\\
%a Dipartimento di Fisica e Astronomia, Univ. of Catania and INFN,
%I-95123 Catania, (Italy) \\
%b Dipartimento di Fisica and INFN, Univ. of L'Aquila, (Italy) \\
%c Dipartimento di Fisica and INFN, Univ. of Roma II, Tor Vergata, Roma (Italy)
}
\presenter{Presenter: J. Rodriguez Martino (julio.rodriguez@roma2.infn.it),\
ita-rodriguezmartino-J-abs1-he15-poster}

\maketitle

%\linenumbers

\begin{abstract}
As more than half of the Fluorescence Detector (FD) of the Auger Observatory is completed, 
data taking is becoming a routine job. It is then necessary to follow strict procedures to 
assure the quality of the data. An overview of the data taking methods is given. The nature 
of the FD background signal is due to  the night sky brightness (stars and planet faint light, 
moonlight, twilight, airglow, zodiacal and artificial light) and to the electronic
background (photomultiplier and electronic noise). The analysis of the fluctuations in the 
FADC signal (variance analysis), directly proportional to the background mean light level, 
performed  for each night of data taking is used to monitor the FD background signal. The 
data quality is analysed using different techniques, described in detail. Examples of trigger 
rates, number of stereo events, dead time due to moonlight, weather or hardware problems are 
given. The analysis comprises several months of data taking, giving an overview of the FD 
capabilities, performance and allowing a systematic study of data and their correlation with 
the environment.

\end{abstract}

%%%%%%%%%%%%%%%%%%%%%%%%%%%%%%%%%%%%%%%%%%%%
\section{Introduction}
%%%%%%%%%%%%%%%%%%%%%%%%%%%%%%%%%%%%%%%%%%%%

The Pierre Auger Observatory studies high energy cosmic rays in the
region of the Greisen-Zatsepin-Kuz'min (GZK) cutoff \cite{nim_auger}.
One of the detection methods used is the collection of air fluorescence
light generated in the Earth atmosphere by extensive airshowers.
Currently there are 18 fluorescence telescopes in operation,
distributed in 3 "eyes", which overlook the entire surface
array. The last eye is foreseen to be completed in early 2006.

Every eye building houses 6 telescopes, each detecting fluorescence light
by means of a camera formed by 440 photomultipliers (PMTs), covering
a field of view of roughly 30 x 30 degrees. Each PMT works
as one pixel to form the complete image of the shower. 

To assure the quality of the FD data, it is necessary to follow
well defined procedures on data taking and on reporting the problems
that might arise. The study of the background associated with
each measurement is important for the
sensitivity that the telescopes can achieve. Statistics on
different parameters associated with the measurements help
identify hardware and software problems or simply calculate
the effective measurement time. This work gives an overview
of the efforts made to achieve these goals.

%%%%%%%%%%%%%%%%%%%%%%%%%%%%%%%%%%%%%%%%%%%%
\section{Data taking procedures}
%%%%%%%%%%%%%%%%%%%%%%%%%%%%%%%%%%%%%%%%%%%%

Every month, during the period around the new moon, a group
of physicists takes care of the FD data taking shift. They
are instructed on how to operate the data aquisition system
and then supervised until they get aquainted with 
the procedure. All the information that the operators might need during the shift
is contained in an internal web page.
There are links to hardware and software manuals, information about the moon fraction,
twilight times and weather. 

Every occurrence during the shift is reported in the FD
electronic e-log, accessible from the FD web page. A template
is provided to unify the way of reporting different events.

%%%%%%%%%%%%%%%%%%%%%%%%%%%%%%%%%%%%%%%%%%%%
\section{Variances analysis of FD background signal}
%%%%%%%%%%%%%%%%%%%%%%%%%%%%%%%%%%%%%%%%%%%%

The contribution to
the background signal comes from stars and planet faint light,
moon light (in the field of view or diffuse), twilight at the start
or the end of the data taking run, lightning, air glow, zodiacal light and man-made light pollution.
The electronic noise comes from the PMTs and the electronics chain.
Thus, the total background signal is the sum of the sky and
electronic background
$$ S_{ADC}^{bckg} = S_{ADC}^{sky} + S_{ADC}^{ele} \qquad $$

and the ADC signal variance is the sum of the sky background variance and
the electronic noise variance

$$ [\sigma^2_{ADC}]^{bckg} = [\sigma^2_{ADC}]^{sky} + [\sigma^2_{ADC}]^{ele} \qquad .$$

The environment background signal can be studied on an event-by-event
basis, looking at pixels not passing the First Level
Trigger (FLT) \cite{nim_auger} or with special background measurements, by
registering the background value every 30 seconds during the
data taking. To estimate the electronic noise value, we used the background files
acquired with closed shutters every night before data acquisition. This
background measurements last for a few minutes sampling the pixels every 5 seconds.

The direct current induced by the background is eliminated by the AC 
coupling of the PMT base.  The noise will be generated
by the fluctuations of the mean value, which is proportional to the
mean number of photons reaching the PMT. This motivates the
present systematic analysis of the variances.

A stand-alone code has been developed to monitor the FD background
signal and the detector status for any data acquisition night for all
mirrors and telescope sites. The program can process the background
signal in the FD data runs event by event and/or the special
background files with closed or open shutters. The output consists
in ROOT \cite{root} files that can be then analyzed off-line. It
can also plot variances, pedestals, thresholds and pixel hit rates
on a nightly basis using an on-line display.

For each event in the FD data runs, 
%(usually one run per night of
%data taking), 
%the algorithm extracts the acquired pixels and their
the algorithm extracts the pixel information, including their
ADC traces. For each trace, the mean value (pedestal) and the
variance are obtained. Only pixels not triggered by FLT 
logic {\it (noise pixels) } are used while for the
special background measurements all pixels in the camera are considered.

\begin{figure}[h]
   \begin{center}

   \includegraphics[height=7.0cm,width=8.5cm]{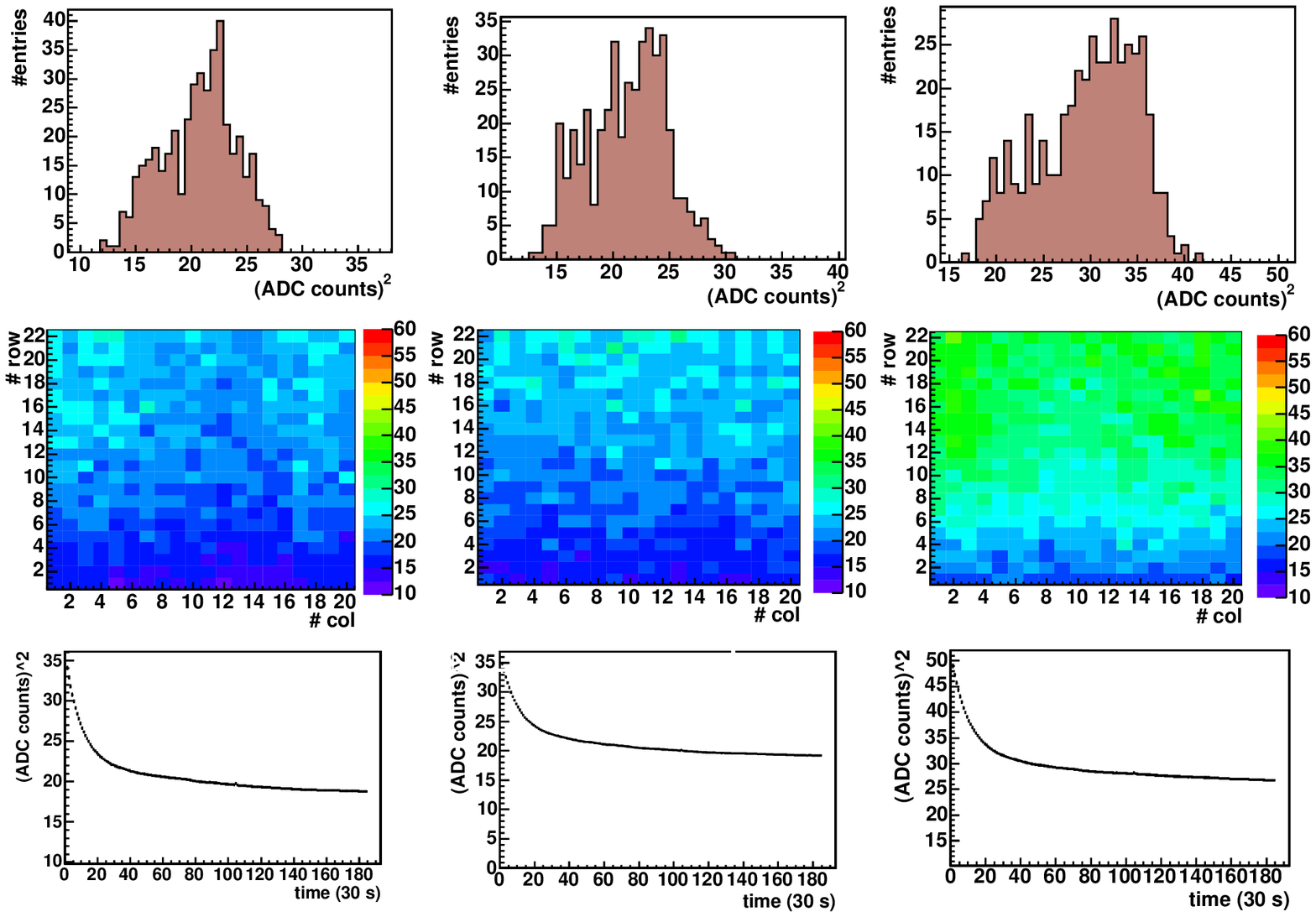}
   \includegraphics[height=6.5cm,width=6.5cm]{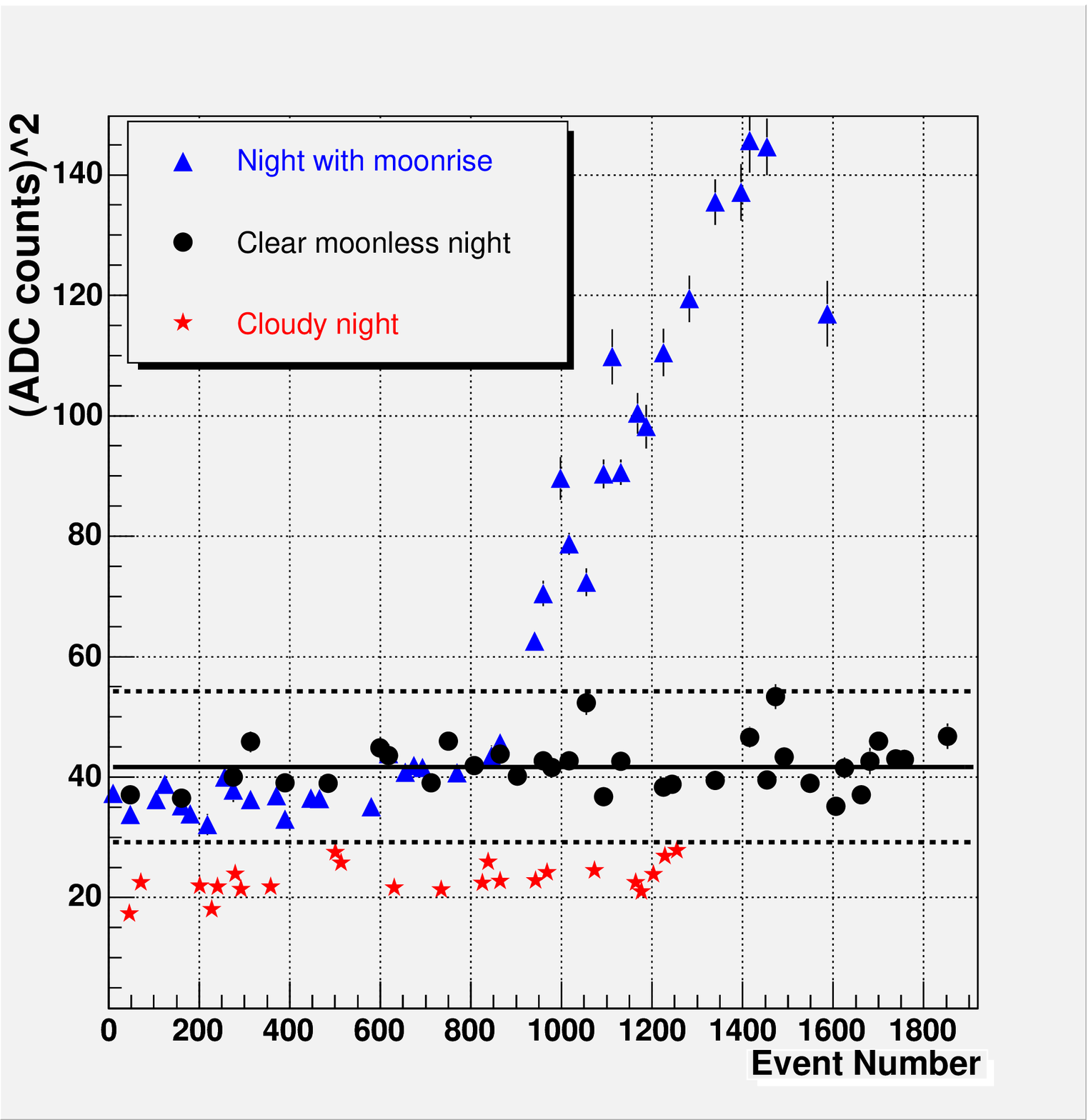}
   
   \caption
   { \label{fig:clearnight} (Left) Total variance (sky background + electronic
noise) for telescopes 1, 2 and 3 in Los Leones eye for a typical clear
moonless night, in terms of
distributions (top) and color maps (middle), 
%and time profiles (bottom). 
%In the distributions and in the color
%maps, 
averaged over the whole night. 
%In color maps
%each pixel is identified by its row and column numbers in y and x
%axes respectively. The side color bars give the value corresponding
%to each parameter. 
Each point in the time profiles (bottom) is the total
variance averaged over all pixels vs. time in 30 s
units. 
%: except for the beginning of data taking  (the presence of
%twilight increases the variances) the behaviour is rather stable
%during the whole night. 
(Right) Variances vs. event number for
   three typical data taking nights: i) (triangles) night with moonrise;
   ii) (circles) clear moonless night; iii) (stars) cloudy night.
   Each point is the average value of the variance taken over all
   noise pixels in telescope 4 at Los Leones. The continuous line is the mean
   value over the whole night for clear moonless conditions, the dashed lines
indicate $\pm 1 \sigma$.}

\end{center}
\end{figure}

%An on-line display is presently available as a tool for the FD operators.
%They can check the variances, pedestals, thresholds and pixel hit rates
%on a nightly basis and make a pre-selection of the data taking runs.
A typical result is shown in Figure \ref{fig:clearnight} (left) where
data from three telescopes in Los Leones eye are shown.

%-------------

%-------------

The results in Figure \ref{fig:clearnight} can help in classifying
the nights according to the data taking conditions, just
looking at the variances.  Typical results are shown in Figure
\ref{fig:clearnight} (right) for telescope 4 in Los Leones. Note that the
variance is very sensitive to the data taking conditions . The
nights can be easily classified, as shown in the figure. The
electronic noise for a typical night is reported in Table I, for all
telescopes in Los Leones and  Coihueco sites. A systematic
investigation for all data taking nights in 2004 has been performed.

%\vfill

%-------------
%   \begin{figure}
%   \begin{center}

%\end{center}
%   \end{figure}

%-------------

\begin{table}[h]
\footnotesize
\begin{center}
\begin{tabular}{|c|} \hline
\textbf{ELECTRONIC NOISE} \\
\hline
\end{tabular} \\
\begin{tabular}{|p{1.4cm}|p{2.6cm}|p{2.6cm}|p{2.6cm}|p{2.6cm}|} \hline
%   \textsf{Los Leones} ~~~~Tel Id  & Variance $  (ADC \, counts)^2$ & Pedestal $  (ADC \, counts)$
%&  Threshold $  (ADC \, counts)$ & Pixel Hit Rate $  (Hz)$\\
%\hline

%1 & $3.5 \pm 0.5$  & $123 \pm 25$ & $1258 \pm 249 $ & $99.8 \pm 2.6$
%\\ \hline

%2 & $3.3 \pm 0.5$  & $119 \pm 26$ & $1217 \pm 265 $ & $99.6 \pm 3.2$
%\\ \hline

%3 & $3.8 \pm 0.8$  & $120 \pm 20$ & $1233 \pm 207 $ & $100.2 \pm
%3.5$
%\\ \hline

%4 & $4.0 \pm 0.9$  & $127 \pm 24$ & $1302 \pm 243 $ & $100.1 \pm
%3.5$
%\\ \hline

%5 & $3.5 \pm 0.8$  & $140 \pm 20$ & $1432 \pm 205 $ & $100.1 \pm
%3.9$
%\\ \hline

%6& $3.7 \pm 0.8$  & $122 \pm 22$ & $1248 \pm 227 $ & $110.5 \pm 3.7$
%\\ \hline

% \hline

   \textsf{Coihueco} ~~~~Tel Id  & Variance $  (ADC \, counts)^2$ & Pedestal $  (ADC \, counts)$
&  Threshold $  (ADC \, counts)$ & Pixel Hit Rate $ (Hz)$\\
\hline

1 & $3.9 \pm 0.3$  & $140 \pm 20$ & $1400 \pm 200 $ & $85 \pm 9$
\\ \hline

2 & $3.8 \pm 0.3$  & $130 \pm 30$ & $1300 \pm 300 $ & $100 \pm 4$
\\ \hline

3 & $4.2 \pm 0.5$  & $120 \pm 20$ & $1200 \pm 200 $ & $100 \pm 3$
\\ \hline

4 & $3.9 \pm 0.6$  & $130 \pm 20$ & $1400 \pm 200 $ & $93 \pm 5$
\\ \hline

5 & $3.8\pm 0.3$  & $140 \pm 20$ & $1400 \pm 200 $ & $86 \pm 8$
\\ \hline

6& $3.8 \pm 0.3$  & $130 \pm 10$ & $1400 \pm 100 $ & $100 \pm 10$
\\ \hline

\end{tabular}

\end{center}

\normalsize

\caption   { \label{fig:elec} Values for variances, pedestals, 
thresholds and pixel hit rates, averaged over all PMTs in
Coihueco. Results obtained from the
electronic noise monitoring with closed shutters.}

\end{table}

%%%%%%%%%%%%%%%%%%%%%%%%%%%%%%%%%%%%%%%%%%%%
\section{Analysis of the data taking period}
%%%%%%%%%%%%%%%%%%%%%%%%%%%%%%%%%%%%%%%%%%%%

After each data taking period, a report is generated with information
about the performance of the FD telescopes. It is divided in a short report 
that shows the measurement and down times for every night, in every FD site, and 
a long report that adds information about every run, start and end times and 
telescope fraction (see section \ref{sec:dead_time}), shows links to PostScript 
and ROOT \cite{root} files containing information about the total number of triggers 
and trigger rate as a function of time. Finally there is a link to the e-log
written during the shift period. 
%The next sections will describe in detail 
%the different parameters analysed and the way of doing this analysis. 
The source of information for this analysis are the log files written by the data aquisition (DAQ)
program %(called eyerc*.log) 
and the data files, which are automatically analysed using a stand-alone program.
% "report\_gen"
%(by J. Rodriguez Martino).

%%%%%%%%%%%%%%%%%%%%%%%%%%%%%%%%%%%%%%%%%%%%
\subsection{Dead time}
%%%%%%%%%%%%%%%%%%%%%%%%%%%%%%%%%%%%%%%%%%%%>
\label{sec:dead_time}

The total available dark time is defined as the time between the end of the astronomical twilight at the 
beginning of the night and the start of the twilight the following day. Everything
that prevents the measurement to be as long as this period is considered dead time.
The main source of this dead time is the presence of the moon close to or inside the 
field of view (FOV) of a telescope. But also bad weather, hardware or software problems can
decrease the measurement time. 

The different types of down (or dead) times are defined as follows:

\begin{list}{*}{}

\item {\bf Moon}: this is calculated using a program written by M. Prouza. The time when the
moon is closer than 5 deg to the FOV of each telescope is used to calculate the period when that telescope is not able to
take data. The operators are requested to close the corresponding shutter at the reported time.

\item {\bf DAQ/Hardware}: it is defined as the time between runs in the same night. Takes care of the
fact that some runs are stopped due to a problem with the hardware or software and a new run is started as
soon as the problem is solved. The assigned time could be modified later if the information in the e-log
shows evidence against the assumption.

\item {\bf Weather/other}: the down time that cannot be classified as any of the previous cases is attributed
to the weather or unknown causes. The real cause is later deduced from the e-log entries.

\end{list}

There is another source of dead time, related to the normal operation of the DAQ software and hardware. As
any other system, it has internal delays that can lead to loss of measurement time. This dead time is
calculated by the system itself. It is not yet taken into account in the reports, but will be in a near
future.

For each run, the measurement time is calculated as the time difference between the start and end times, 
multiplied by the telescope fraction. The telescope fraction is calculated as the number of mirrors used
in the run times 1/6. This is useful when some mirrors are not present in the DAQ (for example, they are taken 
out because the moon is in their field of view). Run measurement times are added to obtain the total 
measurement time of the night. Runs shorter than one minute are discarded. A summary of the obtained
results for one run and the whole night are shown in table \ref{fig:deadtime}. The number of stereo
events is calculated by comparing the GPS times and visually discarding those that could be chance
coincidences between two eyes.

%%%%%%%%%%%%%%%%%%%%%%%%%%%%%%%%%%%%%%%%%%%%
\begin{table}
\footnotesize
\begin{center}
\begin{tabular}{|c|} \hline
\textbf{Los Leones - Night from 2004-12-13  to 2004-12-14} \\
\hline
\end{tabular} \\
\begin{tabular}{|l|} \hline
Run number: 627 \\ \hline
Start time (UTC): 2004-12-14 02:41:54 \\ \hline
End time (UTC): 2004-12-14 07:30:17 \\ \hline 
Total time in this run:  17303 s (04:48:23) \\ \hline \hline
Total dark time:  21300 s (05:55:00) \\ \hline
Total measurement time: 20505 s (05:41:45) 96 \% \\ \hline
Down time due to moon: 0 s (00:00:00) 0.00 \% \\ \hline
Down time due to DAQ: 769 s (00:12:49) 3.61 \% \\ \hline
Down time due to weather/other: 43 s (00:00:43) 0.20 \% \\ \hline
Number of stereo events LL-CO: 6 \\ \hline
\end{tabular}

\caption   { \label{fig:deadtime} Measurement and dead times for one night in 
Los Leones. The number of stereo events registered is also indicated.}

\end{center}
\normalsize

\end{table}

%%%%%%%%%%%%%%%%%%%%%%%%%%%%%%%%%%%%%%%%%%%%

\end{document}